\documentstyle[12pt,aaspp4]{article}

\slugcomment{Accepted for publication in the Astronomical Journal}

\lefthead{Sankrit, Blair, Raymond, Long} \righthead{Cygnus Loop STIS}

\begin{document}

\title{HST-STIS Observations of the Cygnus Loop:  Spatial
Structure of a Non-radiative Shock\altaffilmark{1}}

\author{Ravi Sankrit\altaffilmark{2},
William P. Blair\altaffilmark{2},
John C. Raymond\altaffilmark{3},
Knox S. Long\altaffilmark{4}}

\altaffiltext{1}
{Based on observations with the NASA/ESA \textit{Hubble Space Telescope},
obtained at the Space Telescope Science Institute, which is operated by
AURA, Inc., under NASA contract NAS5-26555.}

\altaffiltext{2}
{Department of Physics and Astronomy, 
The Johns Hopkins University,
3400 N. Charles St., Baltimore, MD 21218}

\altaffiltext{3}
{Harvard-Smithsonian Center for Astrophysics, 
60 Garden St., Cambridge, MA 02138}

\altaffiltext{4}
{Space Telescope Science Institute, 
3700 San Martin Drive, Baltimore, MD 21218}


\begin{abstract}

We present a spatially-resolved ultraviolet spectrum of a non-radiative
shock front in the Cygnus Loop, obtained with the Space Telescope
Imaging Spectrograph (STIS) on board the Hubble Space Telescope (HST).
The spectrum covers the wavelength range 1118~\AA\ - 1716~\AA\ with an
effective spectral resolution of $\sim$12~\AA.  The 0\farcs1 spatial
resolution of these data provides a huge improvement over earlier
ultraviolet spectra, allowing us to study the spatial distribution of
high ionization line emission directly behind the shock front.  We are
able to isolate individual shock features in our spectrum by comparing
the STIS spectrum with a WFPC2-H$\alpha$ image of the region.
Isolating the brightest shock tangency, we identify lines of
N~V~$\lambda$1240, C~IV~$\lambda$1549, He~II~$\lambda$1640,
O~V~$\lambda$1371, O~IV] and Si~IV $\lambda$1400, and
N~IV]~$\lambda$1486, as well as the hydrogen 2-photon continuum.  The
N~V emission peaks $\sim$0\farcs3 behind the C~IV and He~II emission
and is spatially broader.  Also, the observed line ratios of C~IV and
He~II to N~V are higher in our bright shock spectrum than in previous
observations of the same filament obtained through much larger
apertures (and little or no spatial resolution), indicating that there
must be a more widely distributed component of the N~V emission.  We
calculate shock models and show that the observed separation between
the C~IV and N~V emission zones and observed line intensities constrain
the combinations of shock velocity and pre-shock density that are
allowed.

\end{abstract}


 

\section{Introduction}

Faint filaments dominated by Balmer line emission are seen around much
of the periphery of the Cygnus Loop supernova remnant (SNR).  These
filaments are due to shocks driven by the supernova blastwave into
partly neutral material.  The Balmer lines, of which H$\alpha$ is the
strongest, are collisionally excited in a narrow zone behind the shock
front (\cite{che80}).  In the hot post-shock region, different elements
are producing resonance line emission (mainly in the ultraviolet) as
they move to higher stages of ionization.  The shocked gas has not had
time to recombine and cool radiatively and so these shocks are termed
``non-radiative''.  The Balmer lines and the ultraviolet line spectrum
can be used as diagnostics to infer properties of the shock wave and
conditions in the preshock gas (e.g.~\cite{ray91}).

One particular Balmer filament, the brightest in the northeast region
of the Cygnus Loop, has been the subject of detailed study.  Based on
the width of the H$\alpha$ line, and on line strengths obtained from
ground based optical spectra and ultraviolet spectra taken with the
\textit{International Ultraviolet Explorer} (IUE), \cite{ray83}
inferred a preshock density of $\sim$2~cm$^{-3}$ and shock velocities
of 210~km~s$^{-1}$ (for the case of rapid equilibration between ions
and electrons), and 170~km~s$^{-1}$ (for the case of slower Coulomb
equilibration between ions and electrons).  \cite{fes85} used the same
IUE data and optical spectra of a different location along the same
filament and favored a 190~km~s$^{-1}$ shock running into gas with a
density of 1~cm$^{-3}$.  The filament was later observed with the
\textit{Hopkins Ultraviolet Telescope} (HUT) and the far ultraviolet
line strengths were found to be best fit by a 180~km~s$^{-1}$
shock, a preshock density of 2~cm$^{-3}$ and rapid equilibration of
electrons and ions behind the shock (\cite{lon92}).  However, in the
same study, the ratio of line strengths in the O~VI
$\lambda\lambda$1032,1038 doublet indicated significant resonant
scattering along the line of sight and implied a preshock density
between 5~cm$^{-3}$ and 12~cm$^{-3}$.  \cite{hes94} presented an
extensive study of Balmer filaments in the NE Cygnus Loop including new
spectra of the filament studied by RBFG (though with a different slit
orientation).  They found the width of the broad component of the
H$\alpha$ line to be significantly narrower than the value given by
RBFG (130~km~s$^{-1}$ compared to 167~km~s$^{-1}$, FWHM).  This implied
shock velocities of 165~km~s$^{-1}$ and 130~km~s$^{-1}$ for the cases
of rapid and Coulomb equilibration.  To reconcile these values with the
far ultraviolet lines which required higher shock velocities, both HRB
and L92 suggested that the shock may have decelerated rapidly in the
last 200 years or so.

The studies discussed above have established the shock properties in a
broad sense.  They have shown that the brightest Balmer filaments are
due to shocks with velocities between 140~km~s$^{-1}$ and
200~km~s$^{-1}$ running into gas with a density between about
1~cm$^{-3}$ and 10~cm$^{-3}$.  However, refining the limits on the
shock velocity and pre-shock density is crucial for interpreting
ultraviolet spectra since the strengths of several diagnostic lines
change dramatically in this velocity range (\cite{cox85}).
Furthermore, the contradiction in the measured H$\alpha$ widths
measured by RBFG and HRB has not been resolved, and the recent
deceleration history of the shock has not been established.

We take a fresh look at this same Balmer filament using the
\textit{Hubble Space Telescope} (HST).  These observations have the
advantage of spatial resolution over the earlier studies.  It has been
known from theory (e.g.~RBFG) that the Balmer emission comes from a
very narrow region behind the shock, corresponding to at most a tenth
of an arcsecond at the $\sim$~440~pc distance of the Cygnus Loop
(\cite{bla99}).  Therefore the earlier observations represent some kind
of average over the detailed filament structure.  The results from
these observations can be re-examined with a knowledge of the
subarcsecond structure of the filament.  More importantly, the spatial
information itself contains diagnostics for the shock properties.  For
instance, the emissivity of any given line as a function of distance
from the shock front depends on the shock velocity and preshock
density.

Our study consists of a WFPC2 H$\alpha$ image and far ultraviolet STIS
spectra of the filament.  The WFPC2 H$\alpha$ image shows the structure
of the filament in exquisite detail (\cite{bla99}).  The shock wave is
an edge-on sheet, gently undulating along the line of sight, with the
shock front less than 0\farcs1 wide when viewed at exact tangency.  In
this paper we present the first part of our spectroscopic study using
STIS - a far ultraviolet long slit spectrum with the slit placed
\textit{perpendicular} to the shock front.  This orientation minimizes
the effects of curvature of the filaments in the plane of the sky and
allows us a relatively clean look at the spatial structure of the
shock.  The second part (which will be the subject of a future paper)
consists of spectra taken with the slit oriented \textit{parallel} to
the shock front, and stepped through a grid of positions spanning the
width of the filament.


\section{Observations}

The observations were obtained on April 29, 1999 as part of HST GO
program 7289 (PI Blair).  The 52\arcsec~$\times$~0\farcs5 slit was used
along with the G140L grating.  The wavelength coverage is 1118~\AA\ -
1716~\AA, and since the source is diffuse and fills the width of the
slit, the spectral resolution is $\sim$~12~\AA.  Though the slit itself
is longer, the effective aperture length in the spatial direction is
restricted to 25\arcsec\ by the dimensions of the FUV-MAMA detector.
In Figure \ref{SLITPOS} the effective aperture is shown overlaid on a
WFPC2 H$\alpha$ image of the filament.  The slit was placed via blind
offset after first acquiring a nearby star with an optical peak-up
acquisition.  Three exposures with a total integration time of 8202
seconds were taken at this position.  The brightest part of the
filament lying in the slit is at R.A. 20:56:05.80 and DEC.  +31:56:13.9
(J2000).  In the figure the shock is propagating upwards
(i.e.\ approximately NE); the region below the filament is the interior
of the SNR, and the region above is the surrounding interstellar
medium (ISM).

In this paper, we are using standard data products generated on
February 1, 2000 with the ``on the fly'' pipeline calibration available
from STScI.  In this mode, the best available calibration reference
files are used.  The reduced 2-D spectra for each of the three
exposures were co-added.  The Ly-$\alpha$ and O~I airglow lines at
1216~\AA\ and 1304~\AA\ respectively were subtracted from the 2-D
spectrum using line profiles taken along the dispersion axis at a
``background'' ISM location.  Figure \ref{2DFULL} shows the resulting
2-D spectrum with annotations indicating the main spectral and spatial
features.

The slit passes through three distinct shock tangencies (Figure
\ref{SLITPOS}).  We will designate the brightest emission near the
center as the ``bright shock'' and the sharp, narrow filaments above
and below it as the ``north shock'' and ``south shock'' respectively.
The spatial positions of these shocks are marked in Figure
\ref{2DFULL}.  We note that these shocks are in front of and behind the
bright shock only in projection, and physically they are part of the
same ``sheet'' at the edge of the SNR.  Furthermore, the bright shock
itself has some substructure and may include multiple shock crossings
along the line of sight.  N~V $\lambda$1240, O~V $\lambda$1371, C~IV
$\lambda$1549, He~II $\lambda$1640, and the 2-photon continuum from the
bright shock are well detected.  The N~V, C~IV and He~II lines from the
fainter south shock are also detected.  Isolating the emission from the
north shock is more difficult since the emission from these stronger
lines fills the space between the bright and north shocks.


\section{Results}

In Figure \ref{SPEC}, we present the background subtracted spectra of
the bright shock and the south shock.  We obtained the 1-D spectra by
averaging over several rows in the spatial direction, in each case
corresponding to $\sim$ 1\farcs2.  The background spectrum subtracted
from each of these shock spectra was obtained by extracting a 5\farcs6
spatial region ahead of the shock,  smoothing the result by 3 pixels to
lower the noise, and scaling appropriately.  The spectra shown in the
plots have been rebinned over 4 pixels (about 0.6~\AA), and the flux at
wavelengths covering the Ly-$\alpha$ and O~I 1304~\AA\ airglow lines
have been set to zero.  The flux units in the plot are
ergs~s$^{-1}$~cm$^{-2}$~arcsec$^{-2}$~\AA$^{-1}$.  The bright shock
spectrum has been offset (arbitrarily) by 0.15 flux units.  The N~V,
C~IV and He~II lines are seen in both the spectra.  The 2-photon
continuum is visible in the bright shock spectrum, but is much weaker
in the south shock.  Additionally,  weaker emission from O~V, the
O~IV]-Si~IV lines around 1400~\AA\ and N~IV] 1486~\AA\ is detected in
the bright shock spectrum.  Although the signal to noise is low, these
detections are a remarkable advance in spatial sampling compared with
earlier ultraviolet observations (RBFG, HRB, L92).

In Table \ref{TBLFLUX} we present the reddening corrected
fluxes of some emission lines, relative to N~V $\lambda$1240 = 100,
measured in our STIS spectra of the bright and south shocks and in
previous studies of the same filament (RBFG, L92).  Line fluxes in our
spectrum were corrected for reddening following L92, who used $E_{B-V}
= 0.08$ and the mean galactic extinction curve of \cite{sea79}.  We
find from their Table 1 that the observed, uncorrected fluxes of N~V,
O~V, C~IV and He~II have to be multiplied by \textit{2.03, 1.89, 1.82}
and \textit{1.80}, respectively.  However, no account has been taken
of possible resonance line scattering of N~V and C~IV photons.

In Table \ref{TBLFLUX} we see that the C~IV and He~II lines are
significantly stronger, relative to N~V, in our spectra than in the
earlier data.  (Due to the low signal, the O~V flux we measure is very
uncertain and is consistent with L92).  The IUE spectra presented by
RBFG used a 10\arcsec\ by 20\arcsec\ aperture and the HUT spectrum
presented by L92 used a 9\farcs4 by 116\arcsec\ aperture, whereas the
region covered in each of our measurements is approximately 0\farcs5 by
1\farcs2, immediately behind the Balmer shock front.  This is strong
evidence that the line ratios depend upon the spatial scale being
observed.  In particular, the comparison shows that N~V emission must
come from a more extended region than the C~IV and the He~II.  Similar
variations in UV line ratios were seen by HRB in IUE spectra taken at 7
positions along another Balmer filament in the NE Cygnus Loop (their
Filament 2).  For example, their Table 6 shows that the C~IV to
N~V ratio varied between 0.58 and 1.43.  HRB discuss one particular
spectrum where N~V is stronger than expected and explain the enhanced
emission as coming from cooling coronal gas, where the gas was ionized
at an earlier time by a shock with a higher velocity.

The distribution of the He~II, C~IV and N~V emitting gas is important
for understanding the recent evolution of the shock (and thereby the
SNR and surrounding ISM).  However, we will postpone further discussion
of this subject to Paper II, where we will present several long slit
STIS spectra covering a larger spatial region.  Now we turn to the
spatial structure of the emission just behind the shock fronts.

In Figure \ref{SPACE} we present the observed flux in several lines as
a function of position along the slit.  The top panel shows the region
around the bright shock and the bottom panel shows the region around
the south shock.  The fluxes (except for H$\alpha$) have been obtained
from the STIS spectrum by integrating over the lines along each row; in
the case of 2 photon emission, the integration was between
1420~\AA\ and 1460~\AA.  The H$\alpha$ flux is from the WFPC2 image,
taking a cross-cut along the slit, and averaging across its width.  The
counts in the image were converted to flux units using the prescription
given in \cite{hol95}.  The peak of the 2-photon emission is chosen as
the $x~=~0$ position, and $x$ increases towards the interior of the SNR
(i.e.\ shocks move right to left in these plots).  The H$\alpha$
emission profile is aligned so that its peak coincides with the
2-photon emission peak.  In the plots, the profiles obtained from the
STIS spectrum have been binned by 4 pixels (0\farcs1) for the bright
shock (top panel) and by 10 pixels (0\farcs25) for the south shock
(bottom panel).

The separation between the N~V emission zone and the shock front is
clearly resolved in the spatial profiles of the bright shock (Figure
\ref{SPACE}, top panel).  This is expected in non-radiative shocks.
After entering the shock front the gas takes time to reach successively
higher ionization states; and so at a given instant, the ionization
state of the gas increases with distance behind the shock front.
Therefore, H$\alpha$ and 2-photon emission arise immediately behind the
shock front, followed by He~II, C~IV and then N~V emission.  The
\textit{HST} spectrum has enabled us to partially resolve this
stratified structure in a non-radiative SNR shock for the first time.
The separation between the C~IV and N~V emission can also be seen in
the noisier south shock profiles (Figure \ref{SPACE}, bottom panel).
We can quantify the separation between the C~IV and N~V zones by
measuring the locations of the peaks for the bright shock.  For the
noisier south shock, the offset between the leading edges of the C~IV
and N~V zones is a measure of their separation.  We obtain a separation
of 0\farcs30$\pm$0\farcs10 for the bright shock and
0\farcs30$\pm$0\farcs25 for the south shock, and use these values in
the discussion below.  At our assumed distance of the Cygnus Loop (440 pc),
0\farcs3 corresponds to about $2~\times~10^{15}$~cm.


\section{Analysis and Discussion}

The shock structure, including the intensities and spatial profiles of
the ultraviolet lines, depends mainly on the shock velocity and the
pre-shock density.  In the case of a non-radiative shock, such as the
one being considered here, the ionization is not necessarily complete
in the post shock flow, and the resulting spectrum is also a function
of the swept up column density.  Furthermore, the line strengths from
any ionization state of an element heavier than helium scales linearly
with the abundance.  We can conveniently interpret our data and
determine (or constrain) the Balmer filament properties by comparing
the observed spectrum with model predictions.  We use an updated
version of the Raymond shock code (described by \cite{ray79} and
\cite{cox85}) to calculate the shock models.  The code considers a
constant velocity shock running into a uniform medium and calculates
the resulting structure, in particular the line emissivities as a
function of swept up column (or equivalently of distance behind the
shock front).  The most important updates for the present investigation
are the ability to follow electron and ion temperatures separately and
the use of ionization rates based on Coulomb-Born calculations
(e.~g.~\cite{you81}).

In order to systematically examine the dependence of the shock spectrum
on the shock velocity and pre-shock density, we have run a grid of
models varying the shock velocity between 160~km~s$^{-1}$ and
200~km~s$^{-1}$ in steps of 10~km~s$^{-1}$ and with three values of the
pre-shock density: 1~cm$^{-3}$, 2~cm$^{-3}$ and 4~cm$^{-3}$.  We assume
abundances representative of the diffuse ISM, taken from \cite{cow86}.
The He, C, N, O, Ne, Mg, Si, S, Ar, Ca, Fe and Ni abundances are 10.99,
8.40, 7.90, 8.70, 8.09, 7.10, 6.50, 7.51, 6.45, 4.61, 5.80 and 4.26 on
a scale where the logarithm of the H abundance is 12.00.  We are
assuming that no elements are liberated from grains due to the shock
until after the ultraviolet line formation.  This is a valid assumption
since the emission region we are considering is formed within about 200
years after the passage of the shock front whereas studies have shown
that the time scale for grain destruction due to such shocks is a few
thousand years (\cite{jon94}, \cite{van94}).

In addition to these basic quantities, the shock models require other
input parameters.  Following RBFG, the neutral fraction of the
pre-shock gas was taken to be 30\% in all cases.  (The H$\alpha$ and
2-photon intensities depend on the density of neutrals entering the
shock front.  However Ly$\beta$ photons can also be converted to
H$\alpha$ plus 2-photon continuum and contribute significantly to their
intensities.  Therefore using the observed H$\alpha$ and 2-photon
intensities to determine the neutral fraction requires detailed
analysis of the Ly$\beta$ radiative transfer, which is beyond the scope
of this paper).  The extent of equilibration between electrons and ions
in the shock front is also unknown (see e.~g.~\cite{dra93}), but
\cite{gha99} has used Balmer line profiles to show that $T_e$ is at
least 80\% as large as $T_i$ in a slightly faster shock in the northern
Cygnus Loop.  In any case, we find that assuming rapid equilibration in
the shock front or slower Coulomb equilibration in the post-shock gas
has a minor effect on the ultraviolet line strengths and a negligible
effect on the spatial profiles.  (A similar conclusion was reached by
\cite{har99}).  Therefore, in all the models we present, rapid
equilibration ($T_e~=~T_i$ everywhere in the post-shock gas) has been
assumed.  The value of the pre-shock magnetic field is important in
determining the properties of the gas and the grain dynamics in the
cooling zone of the post-shock flow where both the gas and the field
are highly compressed.  However, the structure and emission properties
of the ionization zone that we consider here are independent of the
pre-shock field.  In all the models, we use a value of 0.1~$\mu$G for
the pre-shock magnetic field.  (We find that models using pre-shock
fields of 1~$\mu$G give the same results).  The pre-shock gas
temperature in all models is 10,000~K (e.~g.~HRB).


\subsection{An Illustrative Shock Model}

Before moving on to the main analysis, we present an example of the
emission line structure predicted by a shock model.  The example uses a
shock velocity of 180~km~s$^{-1}$ and a pre-shock density of
2~cm$^{-3}$ which are the ``best fit'' values found by L92, who
compared shock models with their HUT data.

The top panel in Figure \ref{MODS} shows the emissivities of
He~II~$\lambda$1640, C~IV~$\lambda$1549 and N~V~$\lambda$1240,
normalized to their respective maxima, plotted as a function of
distance behind the shock front.  The stratification of the post-shock
gas is evident, with emission from the lowest ionization species
occurring first.  The N~V emission zone is well separated from the He~II
and C~IV zones and extends over a larger spatial region.

In the bottom panel of Figure \ref{MODS} the cumulative intensity (in
ergs~s$^{-1}$~cm$^{-2}$) of the lines is plotted.  The cumulative
intensity of a line at a given distance is the intensity in the line
produced up to that point in the post-shock flow and coming out of the
front of the shock.  We see that the He~II and C~IV intensities level
off about $1.5~\times~10^{15}$~cm and $3.5~\times~10^{15}$~cm behind
the shock front while the N~V intensity continues to rise.  (This is
evident from the emissivity plot where the N~V zone extends beyond
10$^{16}$~cm).  Therefore, for a given shock velocity and pre-shock
density, the emission line spectrum depends upon the spatial extent of
the post-shock region, which in turn depends on the amount of material
swept up by the shock.  In these plots we have shown emission out to
10$^{16}$~cm, here corresponding to a swept up column of about
$2.1~\times~10^{17}$~cm$^{-2}$.


\subsection{Spatial Structure}

We have calculated the separation between the C~IV and N~V emission
peaks for the grid of models, and we plot the results in Figure
\ref{SEP}.  The shock velocity is plotted along the x-axis, and
different symbols are used to represent models with different pre-shock
densities.  The separation between the zones decreases with increasing
shock velocity, and for a given shock velocity the separation scales
linearly with pre-shock density.

The physical separation corresponding to the observed angular
separation of 0\farcs3 depends on the distance of the Cygnus Loop.  We
recently found the distance to the Cygnus Loop to be
$440^{+130}_{-100}$~pc, combining the proper motion of the Balmer
filament (using our WFPC2 image and a 1953 POSS-I image) with the range
of shock velocities found for the filament (\cite{bla99}).  In Figure
\ref{SEP}, the dotted line is at $1.97~\times~10^{15}$~cm, which
corresponds to 0\farcs3 at 440~pc.  The dashed lines are at
$1.02~\times~10^{15}$~cm and $3.41~\times~10^{15}$~cm, which correspond
to 0\farcs2 at 340~pc and 0\farcs4 at 570~pc.  They represent the
limits of the observed separation taking into account the distance
uncertainty as well as the $\pm$0\farcs1 uncertainty in the measurement
for the bright shock (\S3 and Figure \ref{SPACE}).

One of the factors influencing the observed separation is the geometry
of the shock front along the line of sight.  Curvature will typically
have the effect of making the observed separation between the
zones larger than the intrinsic separation.  In order to make a simple
quantitative estimate of this effect, we have assumed that the shock
front can be approximated by the arc of a circle.  We then integrate
the model predicted emissivities along lines of sight that are chords
to the circle.  Using a radius of curvature of $5~\times~10^{17}$~cm
(estimated from the H$\alpha$ image, Figure \ref{SLITPOS}), we find the
following.  For the 180~km~s$^{-1}$ shock with pre-shock density of
2~cm$^{-3}$, the separation is increased to $5.9~\times~10^{15}$~cm and
for the 200~km~s$^{-1}$ shock with pre-shock density of 4~cm$^{-3}$,
the separation is increased to $2.3~\times~10^{15}$~cm.  Therefore,
shocks at the lower end of the velocity range and with lower densities
are ruled out by the observed separation.  However, shocks with higher
velocities and higher densities cannot be ruled out on the basis of these
data.


\subsection{Line Strengths}

As we have seen (\S4.1, Figure \ref{MODS}) the shock spectrum for a
given shock velocity and pre-shock density depends on the amount of
swept up material, or equivalently on the width of the post-shock
region.  In order to compare the grid of models with the observed
spectrum, we need to specify this parameter.  The spectrum in Figure
\ref{SPEC} and the line strengths (from our observations) in Table
\ref{TBLFLUX} are based on a 1\farcs2 region behind the shock front.
This corresponds to about $8~\times~10^{15}$~cm at 440~pc.  We
therefore fix the width of the post-shock region in the shock models to
be this value.  Truncated in this way, the swept up column density
depends most strongly on the pre-shock density (and is largely
independent of the shock velocity).  In our grid of models, these
column densities are approximately $4~\times~10^{16}$~cm$^{-2}$,
$8~\times~10^{16}$~cm$^{-2}$, and $17~\times~10^{16}$~cm$^{-2}$ for
pre-shock densities of 1, 2 and 4~cm$^{-3}$, respectively.

In Figure \ref{INT} we plot the N~V intensities predicted by the
models.  As in Figure \ref{SEP}, the shock velocity is plotted along
the x-axis and different symbols are used for different pre-shock
densities.  The N~V intensity increases with pre-shock density (as is
expected), since a greater column of gas is swept up.  The N~V
intensity decreases with increasing shock velocity.  This happens
because temperatures in the post-shock gas go as the square of the
shock velocity (\cite{ray91}) and in faster shocks, elements are
ionized to higher stages (e.~g.~N~VI for nitrogen) more rapidly.  This
results in fewer collisional excitations that lead to the production of
line photons.

The intensities shown in Figure \ref{INT} apply to the case when the
shock is viewed face-on.  The Cygnus Loop filament however represents
the other extreme - a shock viewed more or less tangentially.
Therefore, in order to compare with the predictions, the observed value
has to be corrected for viewing angle and geometry.  The correction
factor is the ratio of the actual shock area observed to the projected
shock area.  This observed intensity has to be divided by this aspect
ratio to obtain the predicted ``face-on'' intensity.  From a ground
based H$\alpha$ image of the filaments in this region, HRB estimated an
aspect ratio of 10 for the \textit{diffuse} (i.~e.~least tangential)
emission.  From the WFPC2 H$\alpha$ image (presented by \cite{bla99}),
we find that the aspect ratio could easily be as high as 50 for bright
filaments.  The reddening corrected, N~V intensity of the bright shock
is 3.6~$\times~10^{-15}$~erg~s$^{-1}$~cm$^{-2}$~arcsec$^{-2}$ (Table
\ref{TBLFLUX} and \S3).  For an aspect ratio of 50, this corresponds to
a model predicted value of
7.2~$\times~10^{-17}$~erg~s$^{-1}$~cm$^{-2}$~arcsec$^{-2}$, which is
shown as a dotted line in Figure \ref{INT}.

The discussion above shows that the range of shock properties we have
considered predicts the N~V intensity more or less
correctly.  However, our rather crude estimate of the aspect ratio from
the morphology of the filaments in the plane of the sky cannot be used
to place very strong constraints on the shock properties.  For
instance, if the aspect ratio were higher or lower by a factor of 2 for
the observed region, the measured N~V intensity would be compatible
with many of the models.  It would be very difficult, though, to
reconcile the observed value with the n$_0$~=~1~cm$^{-3}$ models  --
the predicted intensity is far too low.  Recalling that the C~IV -- N~V
separation also rules out the low density, low velocity, models (Figure
\ref{SEP}), we conclude that the pre-shock gas ahead of the
non-radiative filament has a density of at least 2~cm$^{-3}$.

In Figure \ref{RAT} we plot the C~IV to N~V line ratio for the
grid of models (truncated at a width of $8~\times~10^{15}$~cm).
The C~IV to N~V ratio is higher for lower values of pre-shock
density because the swept up column density is lower and the
N~V zone is less complete (see Figure \ref{MODS}).  The dependence
of this ratio on shock velocity is more subtle - since
it is affected both by the swept up column and the rate of ionization
to higher ionization states.  The observed C~IV to N~V ratio, corrected
for reddening is 1.01 for the bright shock 0.84 for the south shock.
These values are consistent with the shock models having pre-shock densities
of 2 and 4~cm$^{-3}$.  However, as we discuss below, the intrinsic
ratio of C~IV to N~V can be affected by resonance line scattering.

The edge-on viewing angle implies that we are looking through a
relatively large column of emitting gas.  The optical depth of the N~V
and C~IV lines, which are ground state transitions, can get high enough
to significantly attenuate the line intensities.  To estimate the
effect of resonance line scattering, we have used the emission profiles
from our illustrative model (\S4.1), assumed a single shock front with
a fixed radius of curvature (i.~e.~that it is an arc of a circle), and
calculated optical depths along tangential lines of sight.  We find
that the C~IV optical depth ($\tau_{\rm{C~IV}}$) is invariably higher
than the N~V optical depth ($\tau_{\rm{N~V}}$).  We also find that the
ratio $\tau_{\rm{C~IV}}$/$\tau_{\rm{N~V}}$ increases with increasing
optical depth (of either line).  Specifically, for a radius of
curvature $2~\times~10^{17}$cm, $\tau_{\rm{C~IV}}~=~1.6$ and
$\tau_{\rm{N~V}}~=~1.2$ and for a radius of curvature
$1~\times~10^{18}$cm, $\tau_{\rm{C~IV}}~=~3.7$ and
$\tau_{\rm{N~V}}~=~2.6$.  We note that in calculating the optical
depths we have assumed that the kinetic temperature of the heavy ions
is equal to the proton temperature.  If the ionic temperatures are
higher (as is the case for the remnant SN1006, \cite{ray95}), the lines
will be broader and optical depths correspondingly lower.

The attenuation of the line intensity for a given optical depth,
assuming single scattering, is $(1~-~e^{-\tau})/\tau$.  For the cases
mentioned above, the intrinsic ratio of C~IV to N~V would be higher
than the observed ratio by factors of 1.1 and 1.5 for the smaller and
larger radii of curvature, respectively.  The intrinsic N~V intensity
in these two cases would be 1.7 and 2.8 times the observed (dereddened)
intensity.  In the context of our models, a C~IV to N~V ratio of 1.5
would imply that the pre-shock density is somewhat lower than
2~cm$^{-3}$ (Figure \ref{RAT}).  For this ``high optical depth'' case,
the N~V intensity assuming an aspect ratio of 50, would be about
$2~\times~10^{-16}$~erg~s$^{-1}$~cm$^{-2}$~arcsec$^{-2}$.  (That is,
the dashed line in Figure \ref{INT} would have to be placed much
higher).  Then, to make the observation consistent with a low pre-shock
density, the aspect ratio would have to be over 200 (rather than 50),
which is extreme.  Our models therefore favor a scenario where the N~V
and C~IV optical depths are modest, and the lines are not highly
attenuated.  We note, however, that our treatment is very simplistic.
To make accurate models for the observed emission, a detailed treatment
of the line scattering using a realistic geometry is needed.

The observed value of the He~II to N~V line ratio, corrected for
reddening is 1.36 for the bright shock.  Models predict much lower
values - between about 0.1 and 0.4.  The model does not take into
account the production of the 1640\AA\ line by He~II Ly$\beta$
256\AA\ photons and so a lower predicted flux is expected
(e.~g.~RBFG).  \cite{har99} has calculated the expected
He~II~$\lambda$1640 flux for a range of shock velocities.  From Figure
1 of that paper, we find that our model predictions need to be
corrected by factors of between 2 and 5.  Additionally, any attenuation
of N~V due to resonance scattering would increase the observed He~II to
N~V ratio, leading to a better match between models and
observation.  The models predict O~V to N~V of about 0.08, which is
lower than the observed, reddening corrected value of 0.19 for the
bright shock.  However in this case, the error in the observation is
large enough that this difference may not be an issue.  Again,
resonance scattering of N~V would increase the observed O~V to N~V
ratio.

In conclusion, we point out that there are two kinds of uncertainty
associated with comparing model line strengths with the observations.
First, the observed intensity has to be converted to an intrinsic shock
intensity and the conversion depends on the shock geometry along the
line of sight as well as the effects of resonance line scattering.
Second, the intensities are functions of the elemental abundances, and
of the cutoff column density both of which are input parameters for the
models.  For our chosen ISM abundances, and reasonable geometries, we
find that the observations favor shock velocities $\sim$~170~km~s$^{-1}$
and preshock densities between 2 and 4~cm$^{-3}$.


\section{Concluding Remarks}

We have presented a far-ultraviolet spectrum of a non-radiative
filament on the north-east limb of the Cygnus Loop, obtained with
STIS.  The slit length (and hence the spatial axis of the spectrum) was
chosen to be perpendicular to the filament, and therefore perpendicular
to the shock front.  The brightest lines detected in the 1118\AA\ -
1716\AA\ passband were N~V~$\lambda$1240, C~IV~$\lambda$1549 and
He~II~$\lambda$1640.  The effective spatial resolution of the spectrum,
$\sim$0\farcs1, is an improvement over previous studies of the same
filament and is important because it allows us to separate the emission
from individual shock fronts and also to measure the separation of the
C~IV and N~V emission zones in the post-shock flow.  For a given shock
front, the observed C~IV to N~V and He~II to N~V line ratios in our
STIS spectrum are significantly higher than in spectra taken by IUE and
HUT covering larger areas (several square arcseconds).  This difference
is most probably due to a spatially extended component of N~V emission,
such as cooling gas ionized at an earlier time as suggested by HRB.  We
have compared our spectrum with a grid of shock models and find that
the observed separation of 0\farcs3 between the C~IV and N~V zones
excludes pre-shock densities $\lesssim$~2~cm$^{-3}$ for shock
velocities $\leq$~180~km~s$^{-1}$.  (These models predict much larger
separations between the zones).  Furthermore, models with pre-shock
densities $\lesssim$~2~cm$^{-3}$ predict much weaker N~V and C~IV than
observed (assuming ISM abundances and a reasonable path length of
emitting gas along the line of sight).

As the premier example of a remnant whose optical and ultraviolet
emission is dominated by shocks due to the interaction of the SN blast
wave with the surrounding ISM, the Cygnus Loop plays an important role
in our understanding of ``middle-aged'' SNRs.  The spatial scale of the
emission from the shocks responsible for the brightest Balmer filaments
is a few times 10$^{15}$~cm.  This means (as we have shown) that at the
440~pc distance of the Cygnus Loop, the detailed structure of the
filaments can be resolved by HST.  Observations of the filaments at
this high resolution are critical for a correct understanding of the
shock conditions, since lower resolution observations average over
larger regions which in general contain a complex substructure.  (We
again point out the difference between our and earlier spectra of the
same filament, Table \ref{TBLFLUX}).  On the theoretical side, we need
a realistic 3-dimensional model for the shock geometry and the ability
to compute the effects of resonance line scattering.  This would refine
our interpretation of the observed spectra and place more stringent
limits on the shock conditions responsible for the Cygnus Loop Balmer
filaments.


\acknowledgments

This work has been supported by STScI grant GO-07289.01-96A to the
Johns Hopkins University.


\clearpage

\figcaption{The slit position for the STIS spectrum is shown, 
overlaid on a WFPC2-H$\alpha$ image of the non-radiative filament
(from \cite{bla99}).
The box represents the slit, which is 0\farcs5 wide and has an
effective length (determined by the FUV MAMA detector) of 25\arcsec.
The image uses a logarithmic stretch and is scaled to bring out the
shock structures along the slit.
            \label{SLITPOS}}

\figcaption{The pipeline calibrated 2D spectrum of the non-radiative
shock obtained with STIS.  The 0\farcs5 slit and G140L grating were
used.  Several spectral and spatial features are labeled on the
figure.  The ``bright shock'' corresponds to the brightest H$\alpha$
emission seen in the slit in Figure \protect\ref{SLITPOS}.  (Note that
due to remapping done by the STIS pipeline, the bright shock is
off-center in the 2D spectrum while it is near the slit center in the
image.  Strong airglow lines (Ly$\alpha$ and O~I~$\lambda$1304) have
been subtracted to first order.
            \label{2DFULL}}

\figcaption{Background subtracted spectra of the bright shock and the
south shock.  The spectra were generated by averaging over several rows
in the spatial direction, corresponding to $\sim$1\farcs2.  The
background was obtained by averaging over a 5\farcs6 region outside the
SNR shock and smoothing over 3 pixels.  The spectra shown in the figure have
been rebinned by 4 pixels ($\sim$0.6~\AA) and the flux at wavelengths
covering the Ly$\alpha$ and O~I~1304~\AA\ airglow lines have been set
to zero.  F$_{\lambda}$ has units of
ergs~s$^{-1}$~cm$^{-2}$~arcsec$^{-2}$~\AA$^{-1}$.
            \label{SPEC}}

\figcaption{The observed flux in various lines as a function of position
along the slit for the bright shock (top panel) and the south shock
(bottom panel).  The peak 2-photon and H$\alpha$ emission have been
aligned and chosen to be at $x~=~0$, with $x$ increasing towards the
interior of the SNR (i.e.\ shocks move right to left in these plots).
The flux units are 10$^{-15}$~ergs~s$^{-1}$~cm$^{-2}$~arcsec$^{-2}$ and
normalized as shown.  Note the offset of the N~V profile in both bright
and south shocks.
            \label{SPACE}}

\figcaption{Predicted emission from a shock model with velocity
180~km~s$^{-1}$ and pre-shock density of 2~cm$^{-3}$.  The normalized
emissivity (\textit{top panel}) and cumulative intensity
(\textit{bottom panel}) of the three brightest ultraviolet lines detected in the
STIS spectrum are shown as a function of distance behind the shock.
The intensity plotted in the bottom panel refers to the amount of
emission coming out of the front of the shock and has units of
ergs~s$^{-1}$~cm$^{-2}$.  The angular scale bar of 0\farcs3 assumes a
distance of 440~pc to the Cygnus Loop.  Note how the N~V is produced over
a much broader region than the other lines.
            \label{MODS}}

\figcaption{The plot shows the separation between the peak positions of
the C~IV and N~V emission predicted by a set of shock models.  The
separation is shown as a function of shock velocity for different
values of the pre-shock density (shown by different symbols).  The
dotted line corresponds to the observed separation of 0\farcs3 (Figure
\protect\ref{SPACE}) at 440~pc  The dashed lines indicate the limits of the
separation taking into account the uncertainty in the measured separation
and the uncertainty in the distance.
            \label{SEP}}

\figcaption{The plot shows the N~V intensities predicted by the
models.  The models were truncated at a distance of
$\sim~8~\times~10^{15}$~cm behind the shock front.  As in Figure
\protect\ref{SEP}, the shock velocity is plotted along the x-axis and
different symbols are used for different pre-shock densities.  The
dotted line shows the observed, de-reddened bright shock intensity
divided by 50, to correct for the aspect ratio.
            \label{INT}}

\figcaption{The plot shows the C~IV to N~V line ratio for the models,
truncated (as in Figure \protect\ref{INT}) at
$\sim~8~\times~10^{15}$~cm behind the shock front.
            \label{RAT}}


\clearpage

\begin{deluxetable}{cccccc}
\tablecaption{Observed line strengths
                                   \label{TBLFLUX}}
\tablewidth{0pt}
\tablehead{
  \colhead{Ion} & \colhead{$\lambda$} & \colhead{Bright} & \colhead{South}
                & \colhead{RBFG} & \colhead{L92}
}
\startdata
   N~V  & 1240 & 100 & 100     & 100     & 100  \nl
   O~V  & 1371 &  19 & \nodata & \nodata &  14  \nl
  C~IV  & 1549 & 101 &  84     &  71     &  63  \nl
 He~II  & 1640 & 136 & 154     &  48     &  42  \nl
\enddata
\tablecomments{These are observed values, corrected for reddening.
The corrected N~V intensities for the bright and south shocks are 
3.61 and 1.08~$\times~10^{-15}$~erg~s$^{-1}$~cm$^{-2}$~arcsec$^{-2}$.
}
\end{deluxetable}


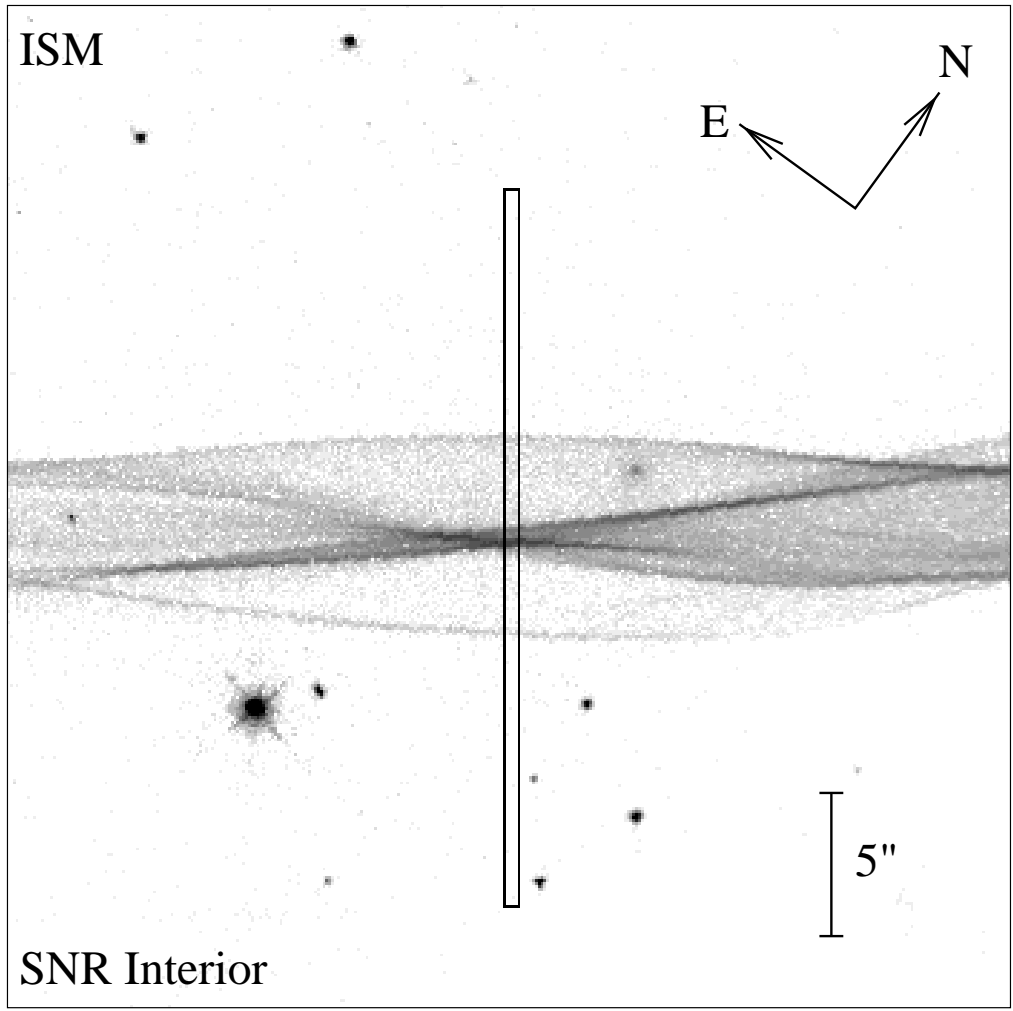
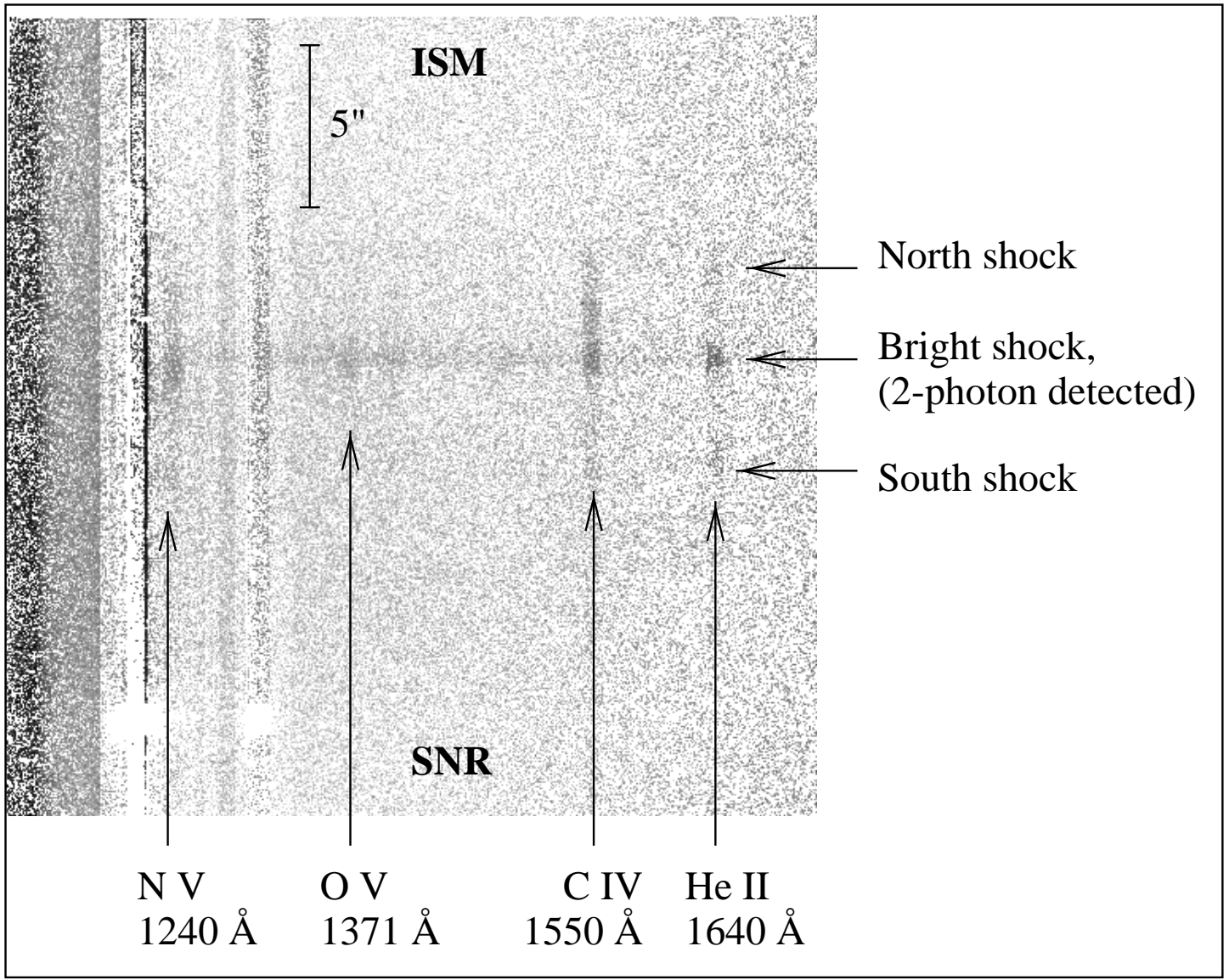
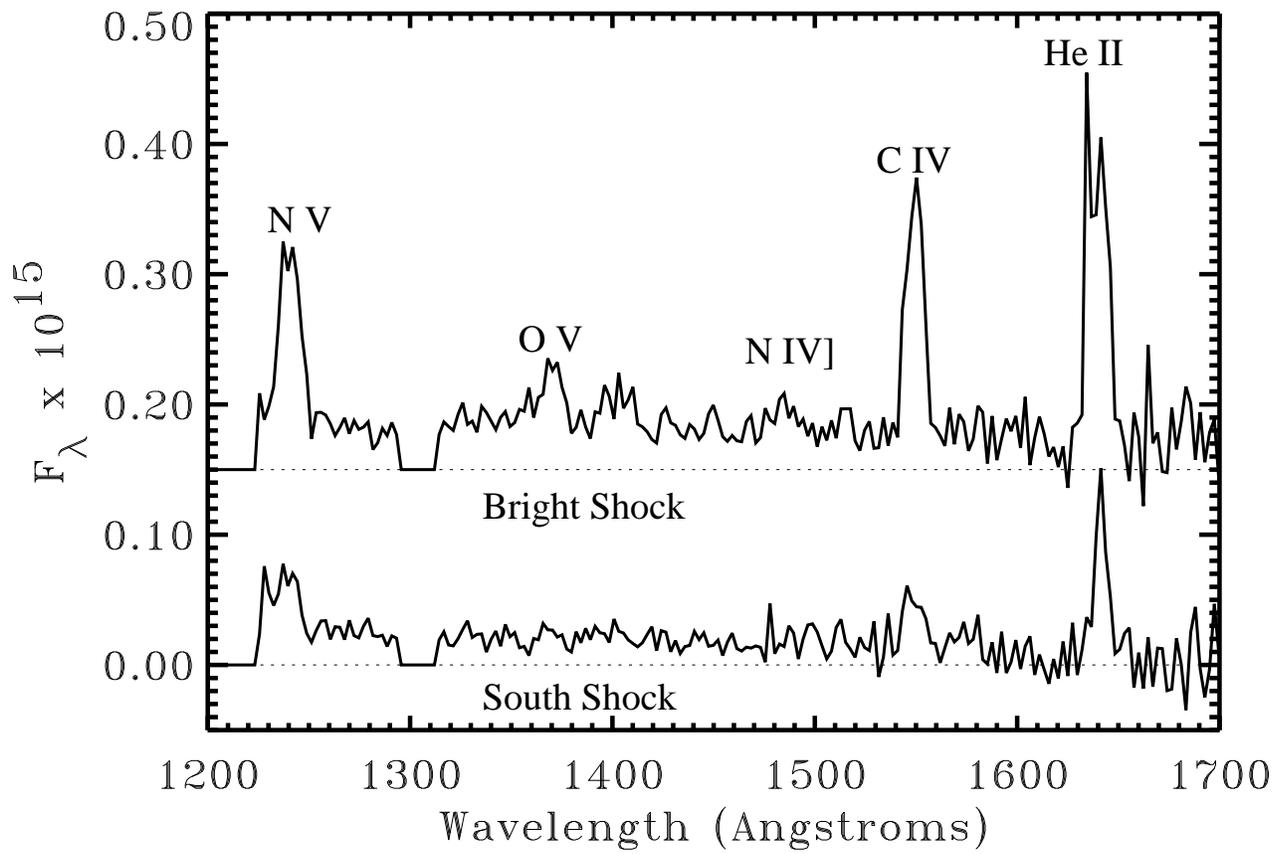
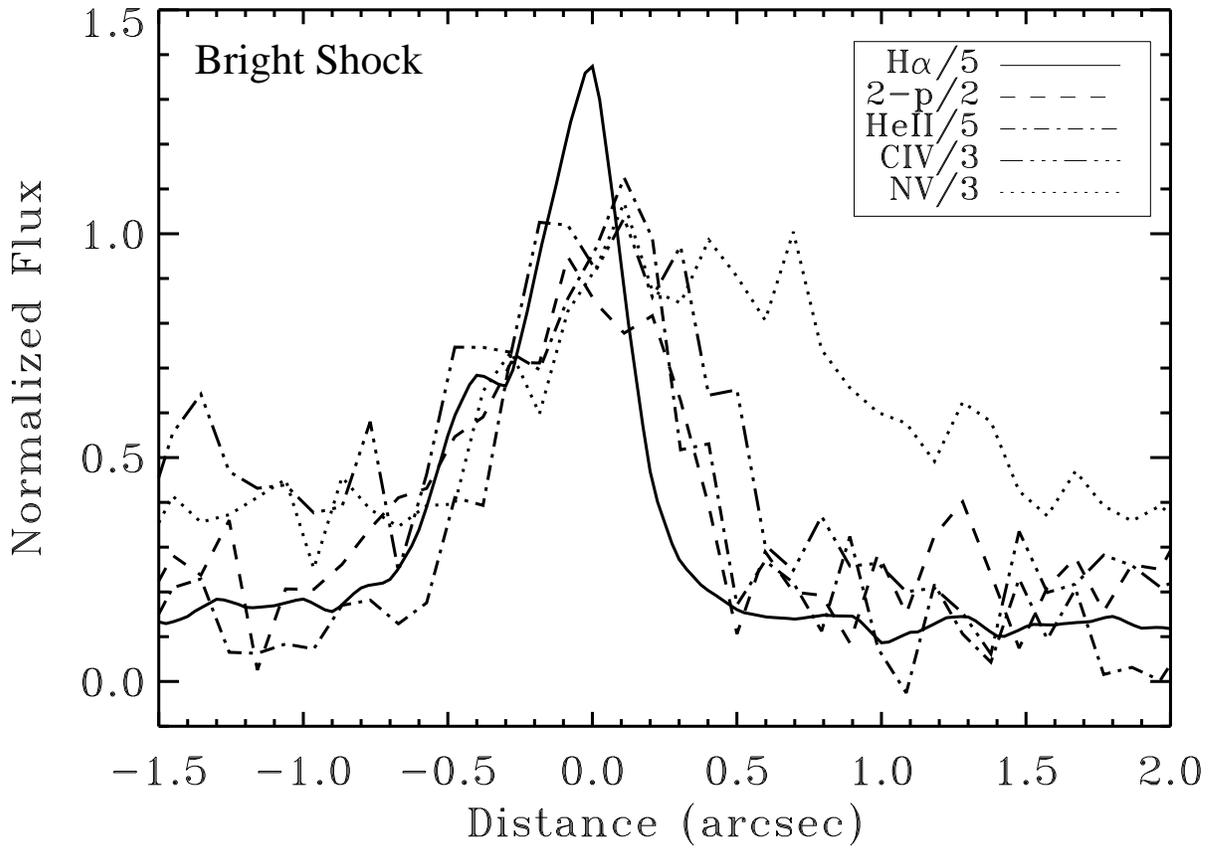
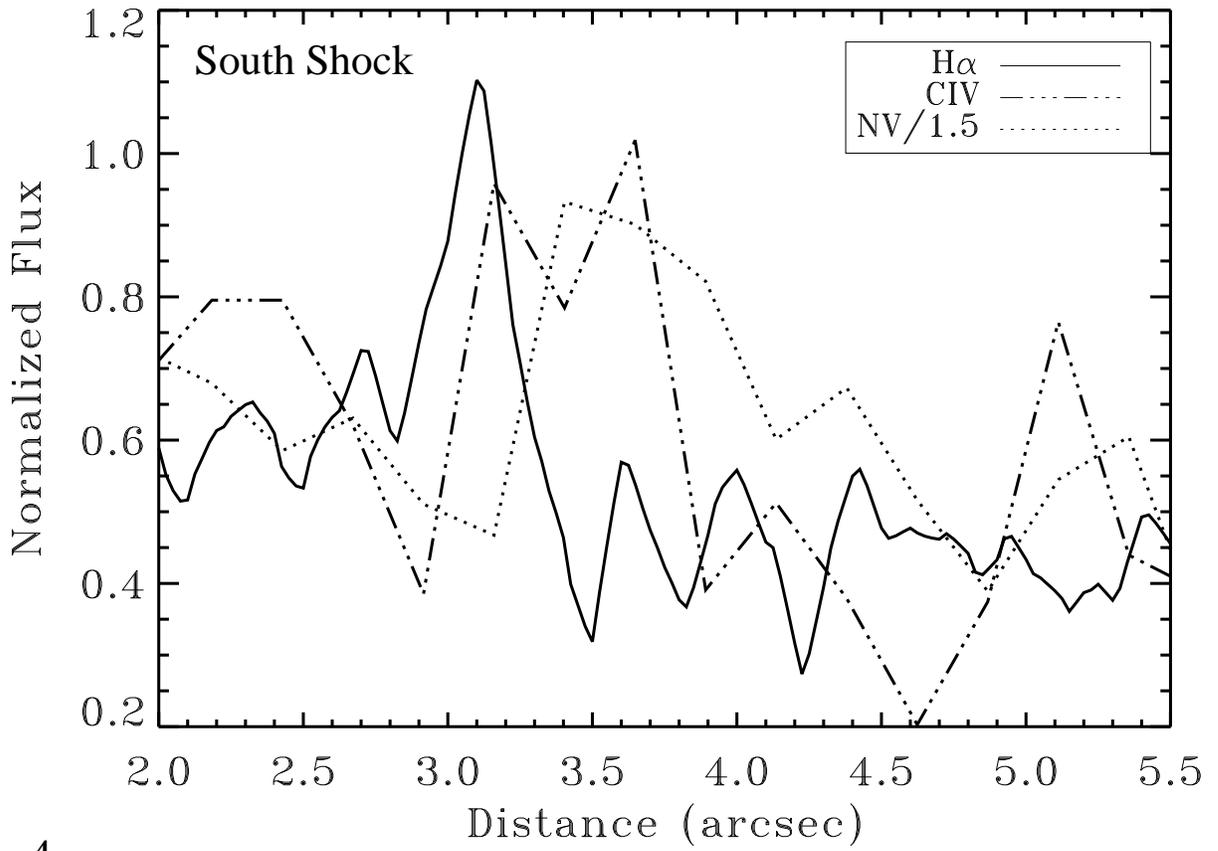
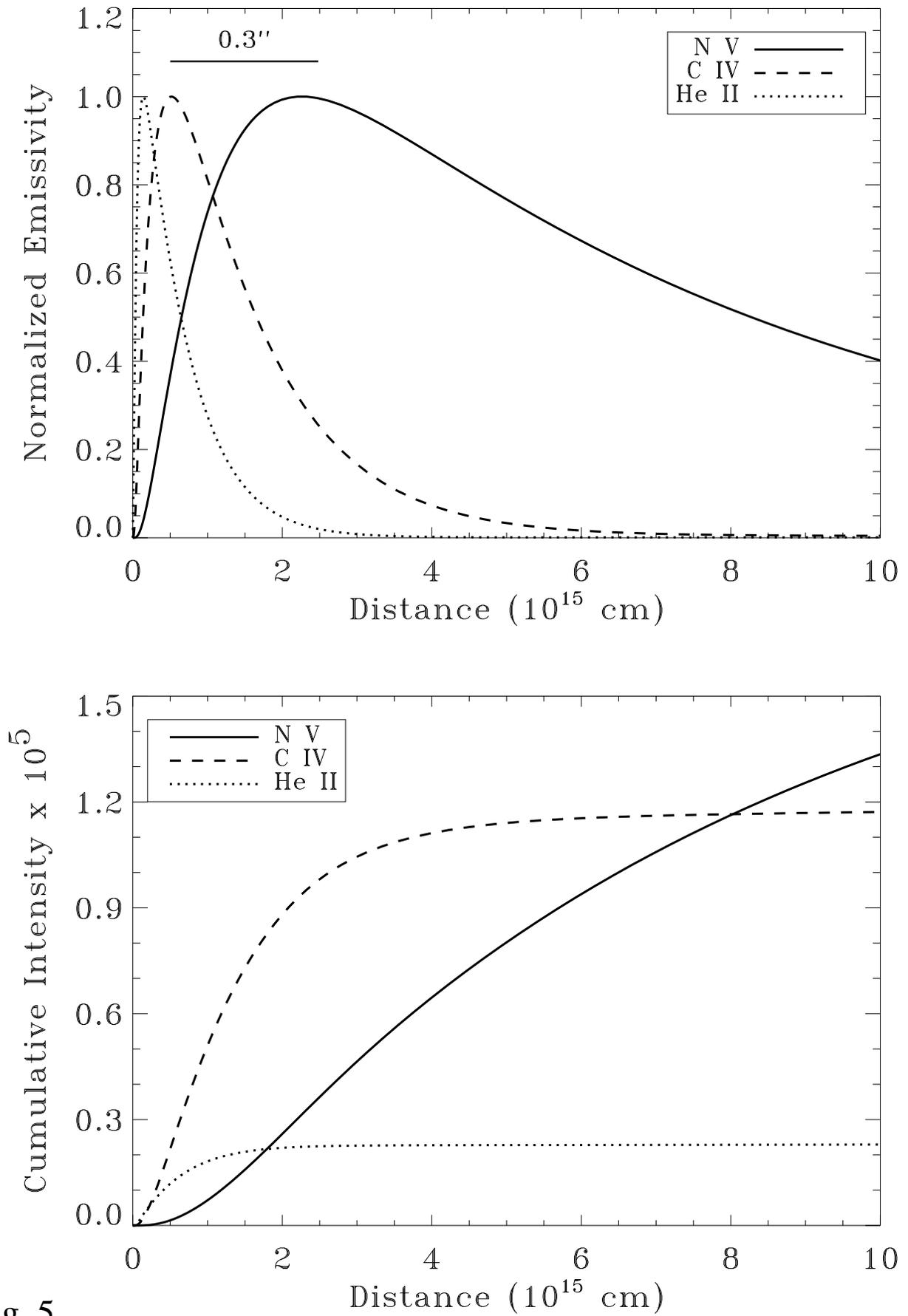
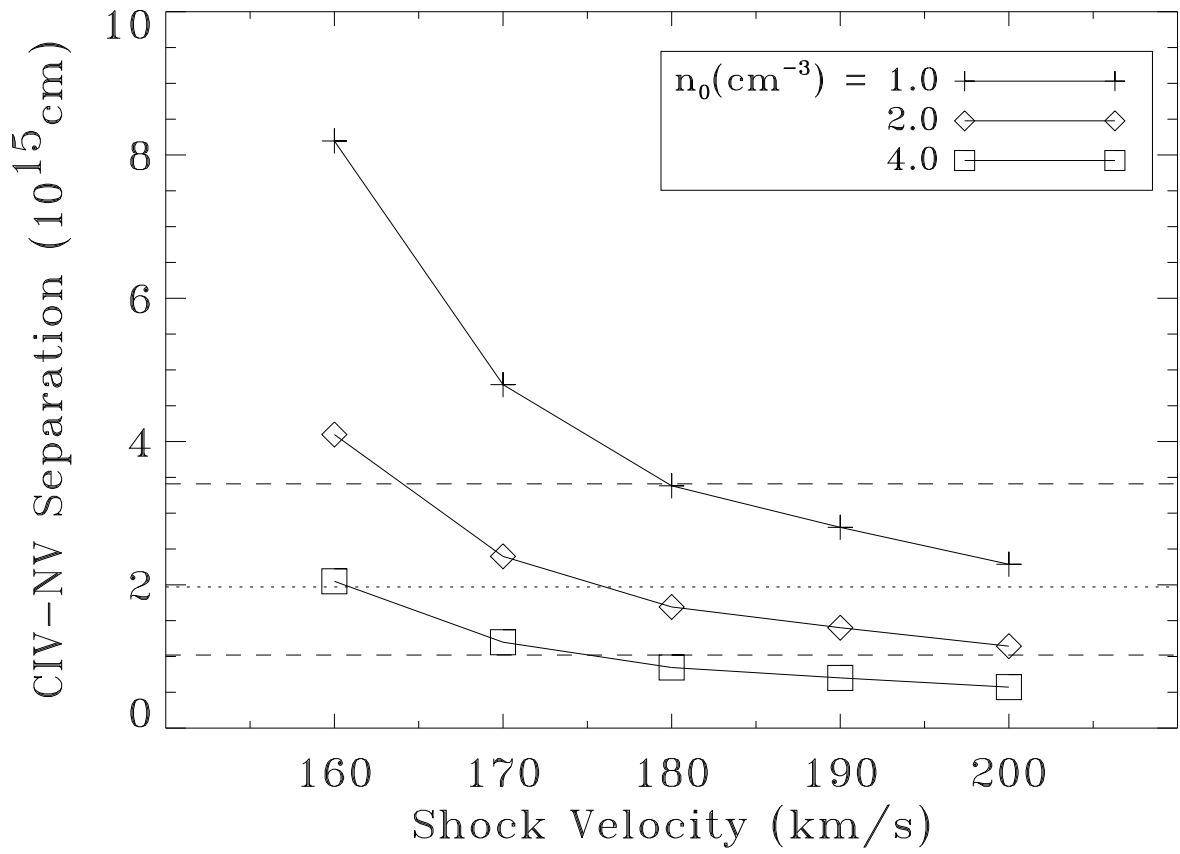
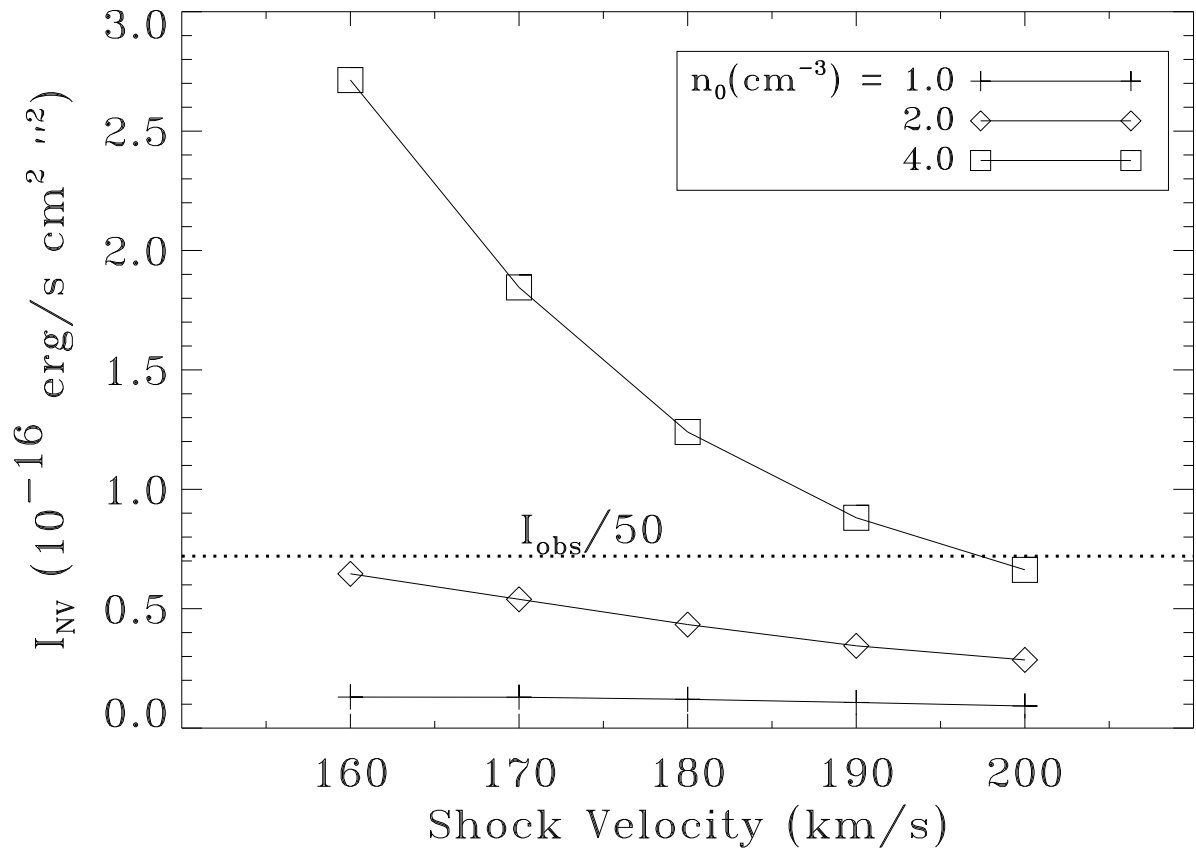
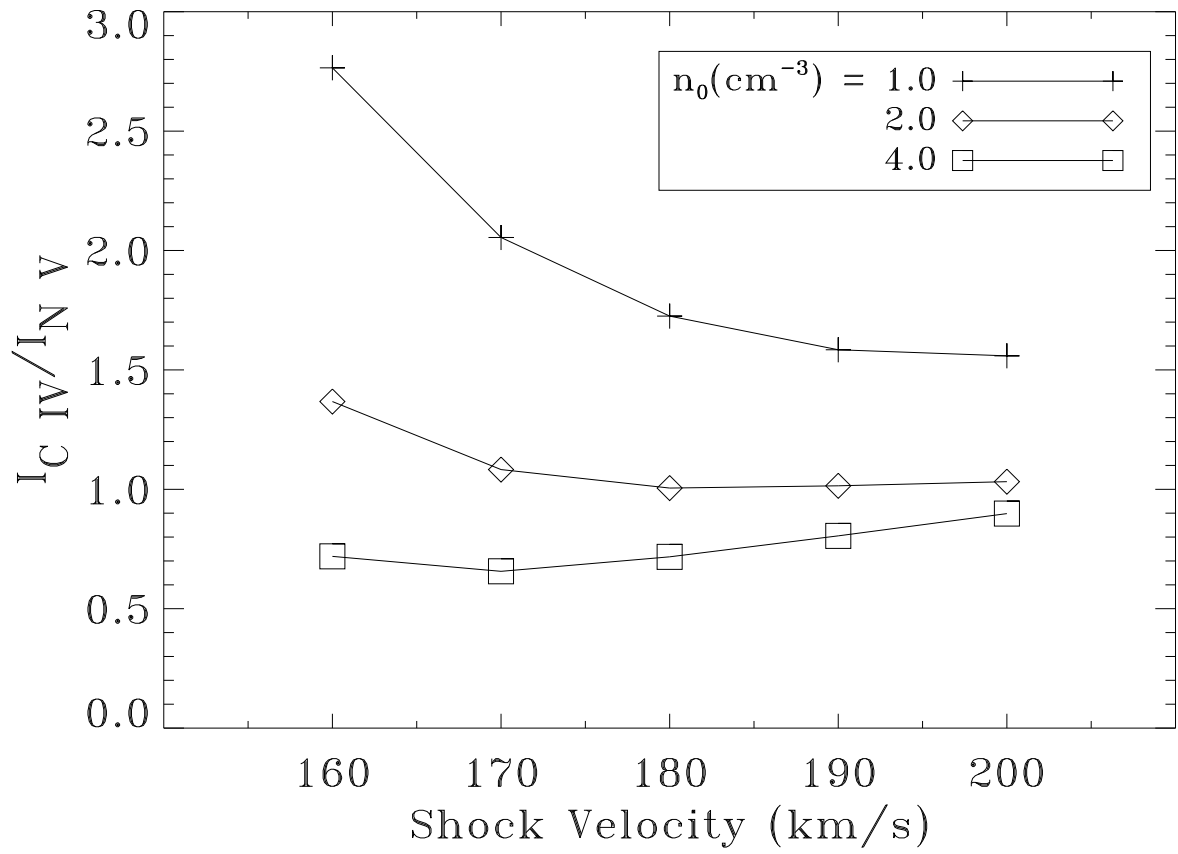
\end{document}